\documentclass[review]{elsarticle}
\usepackage[a4paper, margin=3cm]{geometry}
\usepackage[english]{babel}
\usepackage{graphicx}
\usepackage{setspace}
\usepackage{xcolor}
\usepackage{multirow}
\usepackage{amsmath,amssymb}
\usepackage{array}
\usepackage{lineno}
\usepackage{caption}

\usepackage{numcompress}\biboptions{authoryear}
\DeclareUnicodeCharacter{2212}{-}

\begin{document}
	
	%\maketitle
\title{Quantitative estimates of impact induced crustal erosion during accretion and its influence on the Sm/Nd ratio of the Earth}

\author[mymainaddress,ad2]{L. Allibert\corref{mycorrespondingauthor}}
\cortext[mycorrespondingauthor]{Corresponding author}
\ead{allibert@ipgp.fr}
%\ead[url]{www.elsevier.com}

\author[mymainaddress]{S. Charnoz}
\author[mymainaddress,mylastaddress]{J. Siebert}
\author[mysecondaryaddress]{S.A. Jacobson}
\author[mythirdaddress]{S.N. Raymond}

\address[mymainaddress]{Institut de Physique du Globe de Paris, Universit\'e de Paris, 1 Rue Jussieu, Paris, France}
\address[ad2]{Museum f\"ur Naturkunde Berlin, Leibniz Institute for Evolution and Biodiversity Science, Invalidenstrasse 43, Berlin 10115, Germany}
\address[mysecondaryaddress]{Michigan State University, Earth and Environmental Sciences, 288 Farm Ln, East Lansing, MI 48824, USA}
\address[mythirdaddress]{Laboratoire d'Astrophysique de Bordeaux, All\'ee Geoffroy St Hilaire, Bordeaux, France}
\address[mylastaddress]{Institut Universitaire de France}
	
\begin{frontmatter}
	
\begin{abstract}
%\linenumbers
Dynamical scenarios of terrestrial planets formation involve strong perturbations of the inner part of the solar system by the giant-planets, leading to enhanced impact velocities and subsequent collisional erosion. We quantitatively estimate the effect of collisional erosion on the resulting composition of Earth, and estimate how it may provide information on the dynamical context of its formation. The composition of the Bulk Silicate Earth (BSE, Earth's primitive mantle) for refractory and lithophile elements (RLE) should be strictly chondritic as these elements are not affected by volatile loss nor by core formation. However, an excess in $^{142}$Nd compared to the $^{144}$Nd has been emphasized in terrestrial samples compared to most measurements in chondrites. In that case, the Samarium/Neodymium (Sm/Nd) ratio could be roughly 6\% higher in the BSE than in chondrites, as suggested from the $^{146}$Sm/$^{142}$Nd isotope system \citep{boyet2005142nd}. This proposed chemical offset could be the consequence of preferential collisional erosion of the crust during the late stages of Earth's accretion, leaving a BSE enriched  in Sm due to its lower incompatibility compared to Nd \citep{o2008collisional, boujibar2015cosmochemical, bonsor2015collisional, carter2015compositional, carter2018collisional}. However, if the present $^{142}$Nd of the BSE arises from nucleosynthetic heterogeneities within the protoplanetary disk \citep{burkhardt2016nucleosynthetic, bouvier2016primitive, boyet2018enstatite}, then the BSE has no excess in Sm compared to Nd and this hypothesis precludes any significant loss of relatively Nd-enriched component early in the Solar System. Here, we simulate and quantify the erosion of Earth's crust in the context of Solar System formation scenarios, including the classical model and Grand Tack scenario that invokes  orbital migration of Jupiter during the gaseous disk phase \citep{walsh2011low, raymond2018solar}. We find that collisional erosion of the early crust is unlikely to explain the proposed superchondritic Sm/Nd ratio of the Earth for most simulations. Only Grand Tack simulations in which the last giant impact on Earth  occurred later than 50 million years after the start of Solar System formation can account for this Sm/Nd ratio. This time frame is consistent with current cosmochemical and dynamical estimates of the Moon forming impact \citep{chyba1991terrestrial, walker2009highly, touboul2007late, touboul2009tungsten, touboul2015tungsten, pepin2006xenon, norman2003chronology, nyquist2006feldspathic, boyet2015sm}. However, such a late fractionation in the Sm/Nd ratio is unlikely to be responsible for a 20-ppm $^{142}$Nd excess in terrestrial rocks due to the half life of the radiogenic system. Additionally, such a large and late fractionation in the Sm/Nd ratio would accordingly induce non-observed anomalies in the $^{143}$Nd/$^{144}$Nd ratio. Considering our results, the Grand Tack model with a late Moon-forming impact cannot be easily reconciled with the Nd isotopic Earth contents.
\end{abstract}
\end{frontmatter}


%\linenumbers

\section{introduction}
Chondrites are primitive meteorites that did not differentiate into a core, silicate mantle and crust, and therefore are thought to reflect the chemical compositions of their entire parent bodies. Chondrites have accordingly been considered as the most appropriate proxy for the bulk composition of Earth \citep{ringwood1966chemical, allegre1995chemical, boyet2005142nd}. Whether the Earth building blocks are made of carbonaceous, ordinary or enstatite chondrites, or chondritic at all is still a matter of debate \citep{warren2011stable, dauphas2017isotopic, drake2002determining}. Notably, considering isotopic compositions, Earth is more similar to enstatite chondrites \citep[e.g.][]{javoy1983stable, clayton1984oxygen, cartigny1997nitrogen, dauphas2004cosmic} while considering elemental abundances, carbonaceous chondrites are closest to Earth except for volatile elements \citep{mcdonough1995composition, halliday2013origins}. \newline
Assuming the chondrites represent Earth's building blocks, the bulk silicate Earth should be in chondritic proportions for refractory and lithophile elements (RLE)  as they are neither affected by volatile loss nor core formation (i.e. RLE do not enter metal or sulphide phases). Sm and Nd are considered as RLE that can constitute powerful tracers of collisional erosion because of their geochemical properties (e.g. RLEs, incompatibility degrees, geochronological properties. $^{146}$Sm decays to  $^{142}$Nd with a half life of $\sim$68-103 Myr \citep{kinoshita2012shorter, friedman1966alpha, meissner1987half}. Geochemical studies have enlightened a $^{142}$Nd/$^{144}$Nd ratio for terrestrial rocks of roughly 20 ppm (part per million) higher than various types of chondrites (ranging from $-35\pm15$ppm for carbonaceous chondrites to $-10\pm12$ppm for enstatites) \citep{boyet2005142nd}. This small but significant difference might suggest that the BSE Sm/Nd ratio is $+6\%\ (\pm 1\%)$ above the average chondritic value \citep{boyet2005142nd}. However after 68-103Myr, half of the initial amount of $^{146}$Sm would haved decayed already in $^{142}$Nd, implying only half the effect of a $^{142}$Nd/$^{144}$Nd fractionation onto the Sm/Nd ratio. Additionally, the 6\% excess in Sm/Nd ratio estimate implies that the fractionating process would have occurred within the 30 first millions years of the solar system formation in order to preserve a low fractionation in $^{143}$Nd/$^{144}$Nd \citep{boyet2005142nd}.\newline 
A possible explanation to an excess in the Earth Sm/Nd ratio compared to chondrites \citep{boyet2005142nd} is that the Earth's mantle underwent an early fractionation event induced by partial melting during the first few hundred Myr of solar system history. \cite{mojzsis2019onset} have shown that extensive crustal melting can still happen after 100 Myr due to late accretion. Nd is enriched in magmas with respect to Sm during melting process,  i.e. Nd is more incompatible than Sm \citep[e.g.][]{workman2005major}. Thus, an early crust-mantle differentiation can produce an incompatible-elements depleted mantle corresponding to the present BSE as well as an enriched reservoir with low Sm/Nd corresponding to a primitive basaltic reservoir. This missing  reservoir could be still hidden in the deep mantle after subduction of the primordial crust or formation of a basal magma ocean at the core mantle boundary \citep{boyet2005142nd, campbell2012evidence} that would be unsampled. Another possibility is that this reservoir represents Earth's early crust, which was collisionaly eroded during the late stages of planetary accretion \citep{o2008collisional, campbell2012evidence, boujibar2015cosmochemical, bonsor2015collisional, carter2015compositional, carter2018collisional, shibaike2016excavation}, leading to the inferred superchondritic Sm/Nd ratio. Total erosion of crust about $1\%-1.5$ of Earth mass may be required to account for the fractionation of the BSE Sm/Nd ratio \citep{o2008collisional} if Earth's building blocks were made of either ordinary or carbonaceous chondrites. This number is highly dependant on the partial melting rate assumed in the model \citep{o2008collisional}, on the crustal composition (MORB type here) but also on the dependency with the one-impact erosion assumption. Other estimates can be provided considering other evidences. For example, considering the Mg/Si ratio as well as the hypothesis that the Earth building blocks are made of enstatite chondrites, \cite{boujibar2015cosmochemical} estimate that 15wt\% of crust may have to be preferentially removed during accretion. However, the Earth is only showing a 10-ppm deficit \citep{burkhardt2016nucleosynthetic} in $
^{142}$Nd relatively to the $
^{144}$Nd for enstatite chondrites instead of 20-ppm when considering the average over all chondrites \citep{boyet2005142nd} suggesting a lower loss of crust required to account for this deficit. \newline 
Additionally, recent studies have shown that the existence of nucleosynthetic anomalies in pristine  planetary precursors such as chondrites and Ca-Al rich refractory inclusions (CAIs) could cause the observed variations in the $^{142}$Nd abundances among planetary materials \citep{burkhardt2016nucleosynthetic, bouvier2016primitive}. Moreover, few CAIs and enstatite chondrites have also been shown recently to share common Nd isotope signatures with terrestrial rocks \citep{bouvier2016primitive, boyet2018enstatite}. These observations would imply a chondritic Sm/Nd ratio for the Earth and would preclude dynamical accretion scenarios leading to efficient crustal erosion. However, planetary accretion processes involve energetic impacts with potentially significant erosion of material from embryos \citep{leinhardt2011collisions, leinhardt2015numerically, bonsor2015collisional, marcus2009collisional}. Thus, it is necessary to  quantify  the efficiency of collisional erosion of the primordial crust to address the issue of Earth's accretion scenario and the nature of its building blocks. In previous studies, the efficiency of this process has been investigated with either a simple erosion model \citep{shibaike2016excavation} focusing only on the period of the terminal lunar cataclysm (TLC), the existence of which is in question \citep{boehnke2016illusory, zellner2017cataclysm, morbidelli2018timeline, hartmann2019history, mojzsis2019onset} or neglecting the effect of a population of small planetesimals \citep{carter2018collisional}.  Notably, \cite{shibaike2016excavation} showed that the eventual TLC would not be responsible for removing the entire Hadean crust but most of it. These are estimations based on excavated material, that is defined as the material that is displaced during an impact. This material is not necessarily ejected, so even lower amounts of ejected material can be expected from one bombardment event as suggested by the TLC with a 2$\times10^{23}$g mass of impactors. However, this already raised the high influence that impacts from a population of small bodies may have on crustal stripping.\newline
Recently, a first step toward constraining the effect of collisional erosion on the Sm/Nd ratio of Earth was achieved, in a realistic dynamical context, using sophisticated erosion scaling laws (fitted to SPH simulations of impacts between similarly-sized bodies) coupled to N-body numerical simulations of Earth's accretion \citep{carter2018collisional}. However, computational resources limited the resolution (i.e. number of particles forming planetary bodies) in such simulations: collisions in which the impactor was less than 1\% of the target mass were neglected \citep{carter2018collisional}. As a consequence, SPH numerical simulations do not address the physics of smaller cratering impacts \citep{holsapple2007crater, housen2011ejecta, svetsov2011cratering}, thus hampering quantitative estimates of total eroded material. Planetesimals likely contribute to the majority of impacts, and so collisional erosion may accordingly be driven by small impactors. The effect of these small impacts on the budget of RLE has been neglected in previous works \citep{carter2018collisional}. In addition, \cite{carter2018collisional} only explored the collisional stripping of the early Earth proto-crust during the first stages of planetary growth (between 1 Myr and 23 Myr) \citep{carter2018collisional} while the presence of numerous planetesimals in the early solar system is predicted up to at least 200 Myr.  Only during this period of time and only considering relatively high mass ratio impacts, they already show that a substantial amount of crustal stripping can be reached (6-9\% of the planetary mass). That raises the question on the effects that smaller impacts may have in addition to giant impacts. 

Here we evaluate the hypothesis that crustal erosion from planetesimal impacts can eventually fractionate Earth's Sm/Nd ratio. The evolution of the Sm/Nd ratio is monitored during the accretion of Earth (from 1-3 Myr to 200 Myr) using an approach combining analytical modeling of the cratering process and N-body planet formation simulations. \newline

In section 2 we describe the numerical model as well as the two dynamical  scenarios explored. In section 3 we present our results for both Grand Tack and classical model. Then we summarize our results and discuss them in the context of the collisional erosion hypothesis and nucleosynthetic anomalies hypothesis.
\section{Method}

\subsection{Code description}
We use N-body numerical simluations from \cite{raymond2009building, jacobson2014lunar} to provide the collisional history of terrestrial embryos (whose sizes are defined for each scenario later on) in the proto-planetary disc. These simulations last for about 200 Myr, and have initial conditions representing the state of the solar disk 3-4 Myr after CAIs. Erosion from small impactors (these do not include the impacts fragments that may contribute to the population of small planetesimals) is tracked through the entire accretion process until proto-Earth embryos reach Earth-like mass. 
We explore two scenarios for the giant planet's orbital evolution : (1) the Grand Tack \citep{jacobson2014lunar} and (2) a classical scenario where the orbits of the giant planets are excited by mutual perturbations  \citep{raymond2009building}.  \newline
The impacts are divided in two groups : giant embryo-embryo impacts and planetesimal-embryo impacts. The respective ranges of mass for embryos and planetesimals are dependent on the N-body simulation and are described later on. The planetesimal-embryo impacts are treated with an analytical cratering model based on scaling laws for rocky targets  \citep{svetsov2011cratering, holsapple2007crater}. In our model, giant impacts have no effect on the BSE Sm/Nd ratio as they are assumed to fully melt and mix the crusts and mantles of the colliding embryos. This assumption is a simplification that is made because the melting processes resulting from giant impacts are still poorly constrained. Since evidences have been enlighted by previous studies arguing for only incomplete mantle melting even for very energetic impacts \citep[e.g.][]{nakajima2015melting, carter2020energy}, a first step toward estimating the influence of this hypothesis is made by proposing the three distinct sets of assumptions for the chemical mass balances. The results are presented in the supplementary material. 
After quantifying the mass transfer following each impact, a geochemical model is used to follow the evolution of the amounts of Sm and Nd in the crust and mantle of the different embryos. 
As numerous works suggest an early differentiation of planetary bodies \citep{lugmair1998early, srinivasan199926al, bizzarro2005rapid, amelin2008u, kruijer2014protracted}, we assume that all embryos in the simulations are differentiated. \newline
The processes of crust-mantle reequilibration after an impact are poorly known. Consequently, we assume that giant impacts induce a full crust-mantle reequilibration, while planetesimals impacts only affect the composition and mass of the crust and leave the mantle unchanged. Since the N-body numerical simulations used here assume perfect merging, there is no way of tracking the ejected material. In consequence possible capture of ejected material by other embryos, following an impact, has been neglected. Accordingly, the fractionation computed in this work might be overestimated. We introduce $\epsilon$ the   parameter describing the  chemical fractionation of the Sm/Nd ratio in the mantle of growing Earth like embryos (i.e. BSP for Bulk Silicate Planet) with respect to the initial chondritic composition:

        $$\epsilon=\frac{\left(\frac{Sm}{Nd}\right)_{BSP}^{t}-\left(\frac{Sm}{Nd}\right)_{BSP}^{ini}}{\left(\frac{Sm}{Nd}\right)_{BSP}^{ini}},$$ 
    \newline
with $\left(\frac{Sm}{Nd}\right)_{BSP}^{t}$ the Sm/Nd ratio of the BSP at any time t of the simulation and $\left(\frac{Sm}{Nd}\right)_{BSP}^{ini}$ the Sm/Nd ratio of the BSP at time zero of the simulation (i.e. chondritic value).\newline
The details of the model (i.e. quantification of ejected and accreted masses, chemical mass balances and post-processing N-body simulations of accretion) are presented hereafter.

\subsection{Cratering model}
	   			
 The first step of the work consists in the calculation of the eroded and accreted masses during a single impact event. As we consider embryos as differentiated bodies, the eroded mass is assumed to have a crustal composition when the total ejected mass after considering all impacts from a given distribution is lower than the crustal mass. This assumption is made to take into account the fact that overlapping between the craters may occur since it physically implies that the model assumes that all planetesimals from a given distribution are coming from all direction.  For greater ejected masses, the eroded material is assumed to be made of both crust and mantle (the remnant mass after the entire crust is considered is then ejected mantle). We assume that planetesimals have a bulk chondritic composition which imply that accreted mass has a chondritic composition. 
 The ejected and accreted masses are deduced from previous studies analytical laws constructed as a function of the impact velocity at encounter, the impactor density and the target density and masses of both the target and impactor \citep{holsapple2007crater, shuvalov2009atmospheric, svetsov2011cratering}.\newline

These parameterizations assume that the impactor mass is much smaller than the target mass (i.e. $M_i/M_e < 0.01$) so they cannot be used for giant impacts. The treatment of the giant impacts in the proposed model is dependant on the mixing scenario choosen. This point is described in the section~\ref{sec:mass_balance}.
	
	\subsection{Mass balance model}
	
	\label{sec:mass_balance} \cite{o2008collisional} proposed a mass balance for a two-stage model of crustal erosion: with (1) crust-forming process and (2) collisional erosion. The first stage assumes the formation of the embryo's crust with fractionnation between compatible and incompatible elements. The newly formed proto-crust is enriched in incompatible elements relatively to the residual mantle. In the calculation proposed by \cite{o2008collisional}, $f_{p-c}^1$ represents the fraction of proto-crust with respect to the entire embryo's mass. We adapt this melting equation to our specific problem and define the term $f_{p-c}^1$ as the melt fraction normalized to the primitive mantle mass only, except for the first stage (before any impact) made to estimate the initial composition of the embryo's crust. In the second stage, a $f_{p-c}^2$ fraction of the proto-crust is eroded (note that $f_{p-c}^2<f_{p-c}^1$). The concentrations of an element $M$ in the crust and mantle after the first stage can be described following:
	
	\begin{equation}
		\frac{C_M^{p-c}}{C_M^0}=\frac{1}{D_M+f_{p-c}^1(1-D_M)},
		\label{eq:MBdiff}
	\end{equation}
	With $D_M$ the partition coefficient of M between melt and solid, $C_M^0$ the chondritic composition for $M$, and $C_M^{p-c}$ its concentration in the proto-crust. It is possible to estimate the concentration in the BSP (Bulk Silicate Planet, i.e. crust and mantle of the growing planetary embryo) after the second stage as (assuming no mantle erosion):

	\begin{equation}
		\frac{C_M^{BSP-eroded}}{C_M^0}=\frac{f_{p-c}^1(1-D_M)+D_M-f_{p-c}^2}{(D_M+f_{p-c}^1(1-D_M))(1-f_{p-c}^2)},
		\label{eq:MBerod}
	\end{equation}
	Where $C_M^{BSP-eroded}$ is the concentration of $M$ in the BSP. $f_{p-c}^1$ is initially fixed to a value of 0.026 following \cite{o2008collisional}. It is then assumed that the partial melting rate remains constant over the accretion history (inducing variations in the crust thickness because of the embryo mass increase) but $f_{p-c}^1$ takes a value of 0.037. Three different scenarios are considered for crustal formation after an impact, with notably two of them being end-members (we call them scenario 1 \& 2. ). \newline
	Further results (section~\ref{sec:results}) are presented for the most realistic case only (called scenario 3), however they are all discussed in appendices.
		
		\subsubsection{Scenario 3: "remixing only for giant impacts" model}
		
		This scenario is an intermediate case in between (1) and (2) that respectively refer to two extreme end members. For the scenario 1, all impacts are assumed to melt the entire proto-Earth mantle and to fully re-equilibrate the chemical composition in the BSP. Scenario 2 assumes that none of the impacts produce any mantle remixing. Accordingly, we propose an intermediate case that is much more realistic: scenario (3). Planetesimal-embryo impacts are treated following scenario (2), see eq.~\ref{eq:MBcrust} while Embryo-Embryo impacts are treated following scenario (1)(i.e. full melting of the BSP with formation of a new crust). Accordingly, the composition of the mantle changes each time a giant impact occurs.  All giant impacts are considered as perfect mergers. For simplifications, we assume that this results in a mass of chondritic composition equal to the impactor mass is accreted to the target. \newline
		
    \subsection{Planetesimals size-frequency distribution}
	
	N-body simulations need high computational requirements, so they can not include a realistic amount of planetesimals for the calculations. This is why we have chosen here to distribute the mass of a single planetesimal impactor into a size-frequency distribution of smaller impactors (i.e. each impact from a planetesimal in the data file is assumed to be a serie of impacts with impactors distributed on a size-frequency distribution (SFD)). That serie of impactors is assumed to fall onto the target at the same time. Considering the mass transfer, this assumption implies that the erosion caused by all impactors in the SFD of a given superparticle is applied first. Then accreted mass is added.

	The differential size distribution of impactors follows a power law   :
	
	\begin{equation}
	    \frac{dN_{sfd}(>r)}{dr}=-Kr^{-\alpha},
	\end{equation}
	with $\alpha$ and K standing for positive constants. $N_{sfd}(>r)$ is the cumulative size distribution (the number of bodies with a radius larger than $r$). We get updon integration : 
	\begin{equation}
	    N_{sfd}(> r)=K\left(\frac{r_{max}^{1-\alpha}-r^{1-\alpha}}{1-\alpha}\right)
	\end{equation}
and the total mass of the distribution can be described as
    
    \begin{equation}
        M_T=\int_{r_{min}}^{r_{max}}-\frac{4}{3}\pi r^3 \rho \frac{dN_{sfd}}{dr}dr,
    \end{equation}
	with $\rho$ standing for the mean density of the planetesimals. So replacing $\frac{dN_{sfd}}{dr}$ in the later expression we get, 
	
	\begin{equation}
	    M_T = \int_{r_{min}}^{r_{max}}\frac{4}{3}\pi r^3 \rho K r^{-\alpha}dr.
	\end{equation}
It is then possible to obtain K:
	
	\begin{equation}
	    K=\frac{3M_T}{4\pi\rho}\times \frac{4-\alpha}{r_{max}^{4-\alpha}-r_{min}^{4-\alpha}},
	\end{equation}
	where $M_T$ (the total mass of the distribution) equals the mass of the impactor given in the N-body simulation. $r_{min}$ and $r_{max}$ have to be fixed. They are arbitrarily chosen as 8m and 800km respectively. The number of bodies in each mass bin is evaluated using the cumulative mass distribution, rounded to an integer number using a random number generator. Accordingly, 800km is taken as the upper bound of this distribution to be slighlty lower than the typical size of a planetesimal in the simulations \citep{raymond2009building, jacobson2014lunar}. It may be noted that a 800km-sized body is likely differentiated. However, whether the planetesimals are differentiated or not makes no difference in our model since we only investigate here the evolution of an elemental ratio of two RLEs. However the change in the minimum radius of a body within the SFD is not critical on the model outcomes (less than 1\% change in the fractionation when choosing minimum radius values of few kilometers). 
	The physical characteristics of the planetesimals are as follows :
	
	\begin{itemize}
	    \item The mean density of the target, $\rho_m$, is assumed to be $5200\ kg/m^3$.
	    \item The crust density, $\rho_t$, is assumed to be $2900\ kg/m^3$.
	    \item The density of impactors (chondritic material), $\rho_{imp}$, is assumed to be $2600\ kg/m^3$.
	    \item The impact velocity is assumed to be: $V_{imp}=\sqrt{V_{\inf}^2+v_{esc}^2}$ with $V_{\inf}$ the velocity of the impactor at infinity (beyond the gravitational attraction of embryos), which is here assumed to be the same for all impactors in the same impact event. This $V_{\inf}$ is deduced from the data of the N-body simulation.
	    \item The radii of target and impactors are calculated using their mean densities and respective masses.
	\end{itemize}
	In this study, the slope of the SFD ($\alpha$) is set at $3.5$ which corresponds to the actual value of the distribution of bodies in the asteroid belt. However, the influence of the choice for $\alpha$ needs to be examined, this is the subject of the section~\ref{sec_an:slope} in the appendices. 
	
	\subsection{Dynamical Scenarios }
	
	The impact model presented in the previous section is applied to two dynamical accretion scenarios of the terrestrial planets accretion: (1) Grand Tack or (2) classical. Within each dynamical model, a variety of parameters are explored such as (i) variable eccentricities and semi-major axis of Jupiter and Saturn (for classical model), (ii) different surface densities of the disc (classical) and (iii) different initial embryo masses (Grand Tack). These different dynamical scenarios are summarized in Table 1 (Classical) and Table 2 (Grand Tack) with their associated results for our model in terms of fractionation in Sm and Nd. 
	
	The classical model \citep[e.g.][]{wetherill1978accumulation, wetherill1985occurrence, wetherill1996formation,kokubo2000formation, chambers2001planets, raymond2009building} is the reference model concerning terrestrial planets formation, however it is well known now that it fails in reproducing several key features of our present day solar system. Other models have been further proposed to better reproduce the structure of the Solar System at its present state. Notably, the Grand Tack model is very popular as it among the first models proposed that can explain the small masses of Mars and the asteroid belt. It also reproduces the excited orbits of the bodies in the main asteroid belt, as well as the mixing between inner and outer solar system objects. 
	
	\subsubsection{Classical scenario}
	    
	    The main assumption considered in the Classical model is that the formation of the terrestrial planets can be treated independently compared to the giant gaseous planets \citep{wetherill1990formation, wetherill1992alternative, chambers2001planets, raymond2006high, raymond2009building, raymond2014terrestrial}. The initial conditions thus assume that Jupiter and Saturn are already formed and placed on fixed orbits. Then, it assumes that the terrestrial planets formed essentially from material originated close from their current position within the disk. Such a growth dynamics induces similar sizes for bodies on adjacent orbits (Kokubo and Ida, 2002). Classical models fail to reproduce the size of Mars.\\ 
	    In our study all simulations start with a disk (from 0.5AU to 4.5AU) of embryos and planetesimals with Jupiter and Saturn already formed. They contain 85-90 embryos and 1000-2000 planetesimals \citep{raymond2009building}. The total mass of the system is equally distributed between planetesimals and embryos. However, eight different sets of initial conditions are explored \citep{raymond2009building}. They refer to different orbits of Jupiter and Saturn and different disk surface densities (See Table 1 for details). Depending on the initial conditions, the initial mass of the embryos in the disk is comprised between 0.005 and 0.1 terrestrial masses.
	
	\subsubsection{Grand Tack scenario}
		
		The grand tack scenario was designed to explain the small size of Mars as this was one of the major issues with the classical accretion model \citep{walsh2011low, raymond2014grand}. The inward migration of Jupiter, followed by Saturn before they both migrate outward - when Jupiter gets caught in the 2:3 resonance with Saturn \citep{masset2001reversing, morbidelli2007dynamics, pierens2011two}-  has been proposed as a mechanism to deplete and excite the Mars region leading to a small size of Mars and a distribution of orbital elements in the asteroid belt compatible with current observations. In the simulations we use here, the system contains a population of embryos and two different populations of planetesimals (i.e. internal planetesimals - within 3AU - and external smaller planetesimals beyond 3 AU) \citep{jacobson2014lunar}. Each simulation begins with approximately 100 embryos and 2000 planetesimals. The N-body numerical simulations data used here consider the system is equally distributed between the different planetesimals and embryos masses \citep{jacobson2014lunar}. However, the initial typical mass of an embryo before the onset of Jupiter's inward migration is a free parameter. Three different values are tested here: 0.025$M_{\oplus}$, 0.05$M_{\oplus}$ or 0.08$M_{\oplus}$ (referenced respectively as 0.025, 0.05 and 0.08 in Table 2). The corresponding average $\epsilon$ values and their $1\sigma$ uncertainty are detailed in Table 2.  

\section{Results and discussion}

\begin{figure}
		\centerline{\includegraphics[width=1.\linewidth]{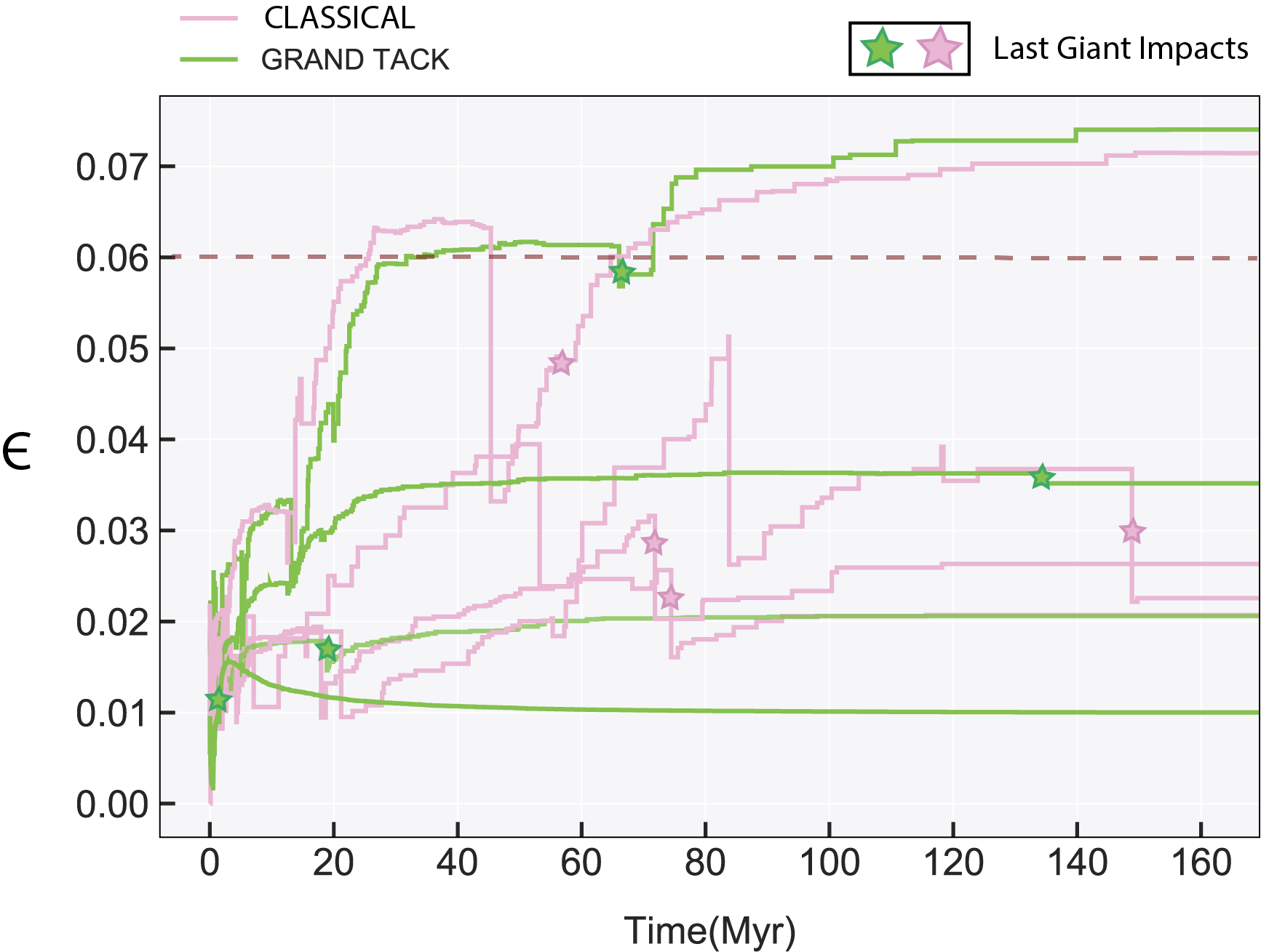}} % this command will be ignored
		\caption{Evolution of the fractionation parameter $\epsilon$ as a function of time for 8 relevant Earth analogs accretion histories: four evolving within a classical accretion scenario with different orbits of Jupiter and Saturn and four evolving within a Grand Tack scenario. They respectively end up with masses of 0.6$M_{\oplus}$, 0.69$M_{\oplus}$, 1$M_{\oplus}$, 1.01$M_{\oplus}$, 1.2$M_{\oplus}$, 1.2$M_{\oplus}$, 1.01$M_{\oplus}$, 0.86$M_{\oplus}$ and semi major axis of 1.16 AU, 0.6 AU, 0.76 AU, 0.62 AU, 0.72 AU, 0.9 AU, 0.95 AU and 0.71 AU, from the most fractionated to the less fractionated. A zero fractionation value refers to a chondritic composition. Time zero is the beginning of the simulation: $\simeq$ 3 Myr after CAIs. In the case of the Grand Tack scenario, time zero corresponds to the very beginning of the Jupiter's migration. Stars refer to the last giant impact for Grand Tack and classical scenarios.}
		\label{fig:epsilon_evol_fig1}
	\end{figure}

\begin{figure}
		\centerline{\includegraphics[width=.7\linewidth]{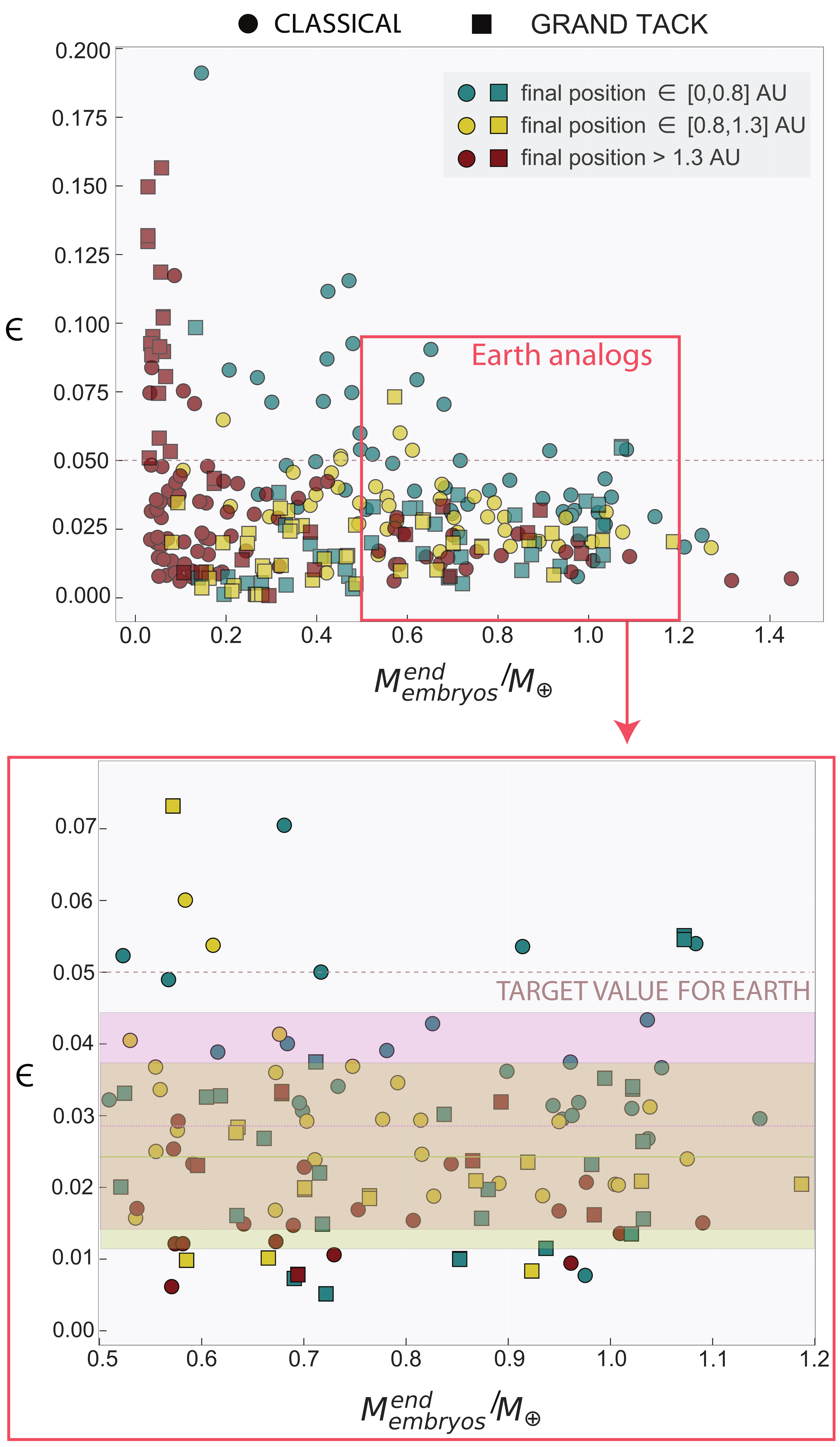}}
		\caption{Top panel: Sm-Nd fractionation with respect to chondritic Sm/Nd values of the different embryos surviving at the end of 40 classical and 31 Grand Tack N-body simulations reported as a function of their final masses. The markers refer to the type of accretionary scenario (circles for the classical scenario and square for grand tack). The colors refer to the final semi-major axis of the embryos. The bottom panel is a zoom of the top panel corresponding to the good Earth analogs. The green line is the average value of $\epsilon$ for the grand tack simulations, the associated light green area is the 1$\sigma$ uncertainty. The pink line and light pink area refer respectively to the average and 1 $\sigma$ uncertainty for the classical accretionary scenario.
		}
		\label{fig:discussion}
	\end{figure}

\begin{figure}
		\centerline{\includegraphics[width=1.\linewidth]{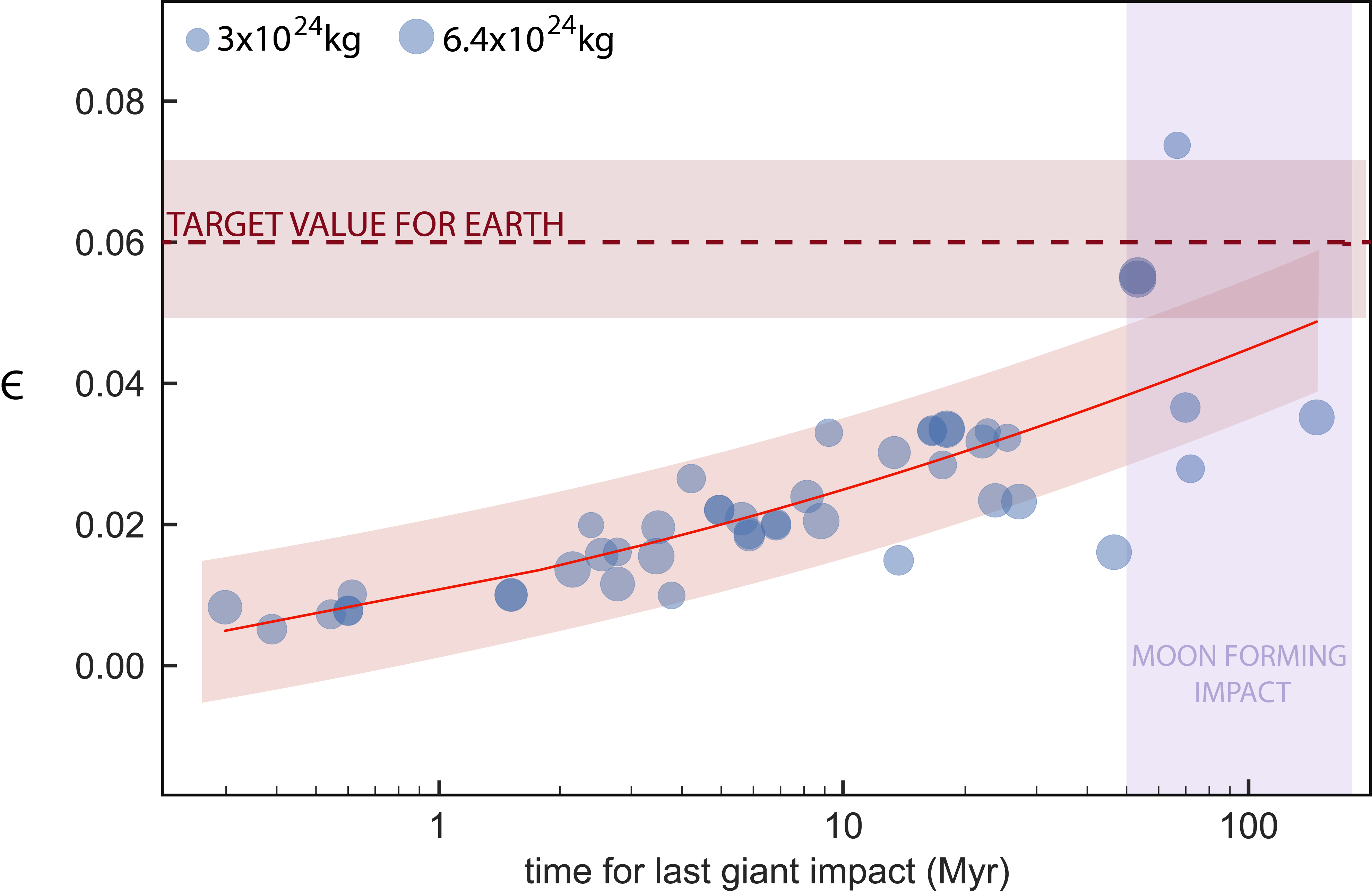}}
		\caption{Sm-Nd fractionation with respect to chondritic Sm/Nd values of the different Earth analogs at the end of Grand Tack N-body simulations reported as a function of the timing of the last giant impact suffered. The markers sizes refer to the final mass of an embryo. The purple area refers to the time laps of the moon forming impact. The red curve is our quadratic fit for the relation between the fractionation value and the timing of the last giant impact, the light red envelop defines the 1$\sigma$ error associated.
		}
		\label{fig:discussion2}
	\end{figure}

Fig.~\ref{fig:epsilon_evol_fig1} shows \label{sec:results} how $\epsilon$ evolves during the accretion of eight representative planets extracted from different N-body simulations: (i) four embryos evolved within a Grand Tack scenario with different initial masses and (ii) four other embryos evolved within a classical scenario for different Jupiter and Saturn orbits. Only a small fraction of the impacts have a significant influence on the fractionation. This effect is a consequence of the impact configuration, especially the impact velocity: the higher the velocity, the more crust is typically ejected. However, successive impacts by planetesimals produce a fractionation that cannot be neglected. The less massive the embryo the easier it is for impactors to erode the surface, leading to a higher fractionation in Sm and Nd in the BSP. The impact history varies considerably from planet to planet and from simulation to simulation. In the case of the Grand Tack scenario, most of the fractionation is produced during the first Myrs, when the system is dynamically excited by the migration of the giant planets. In the classical scenario, the fractionation occurs over the entire accretion history and the final BSP concentration becomes stable later than for the Grand Tack scenario. \newline 
On average neither the classical accretion model ($\epsilon=0.029\pm0.015$) nor the Grand Tack ($\epsilon=0.024\pm0.013$) can account for the eventual superchondritic BSE Sm/Nd ratio ($\epsilon=0.06$), especially not within the first 30 Myrs of accretion. The final $\epsilon$ values are presented in the fig.~\ref{fig:discussion} as a function of the final masses of embryos at the end of the 71 N-body numerical simulations. Most of the surviving embryo's masses range between $0.03M_{\oplus}$ and $1.4M_{\oplus}$. Smaller final embryos present a much larger range of $\epsilon$ values, especially in the case of the Grand Tack scenario (fig.~\ref{fig:discussion}). This is due to the fact that collisional erosion and consequent chemical fractionation is more efficient for smaller embryos. For the Grand Tack scenario especially, the system is very excited during the first Myrs because of the early migration of Saturn and Jupiter. At this time, the embryos can easily be eroded because of their lower masses and escape velocities. However, for bodies considered as relevant Earth analogs (i.e. with a final mass between $1/2M_{\oplus}$ and $1.2\times M_{\oplus}$ and final semi-major axis between 0.5 AU and 2 AU), the range of variation of $\epsilon$ values is narrower than for smaller bodies and $\epsilon$ almost never reaches the Earth target value of 0.06. \newline
Note however that we have used a partial melting rate of 2.6\% as proposed by \cite{o2008collisional} to produce the newly formed crust. Such a low partial melting rate is maybe appropriate for the Earth \citep{o2008collisional} whereas it differs significantly from estimates of differentiated bodies such as Vesta (possibly up to 20\%), Ceres or the Moon (between 5\% to 10\%) \citep{yamaguchi1997metamorphic, ruzicka1997vesta}. However, increasing the partial melting rate in our model would lead only to a decrease in the fractionation of the Sm/Nd ratio (i.e;. lower epsilon value) and could accordingly never account for the observed Sm/Nd value of the BSE. Thus, collisional erosion seems unlikely to explain the possible superchondritic Sm/Nd ratio of the BSE except in marginal cases discussed hereafter. Morevover, the present model assumes striclty chondritic composition for the accreted material which may be invalid especially in the case of  embryo-embryo impacts. A higher fractionation could be expected without this assumption because the impacting embryo might have been fractionated in their own earlier accretion history. Considering the present results, estimating the accurate influence of this parameter cannot be done. This will be the object of further investigation in future works. Additionally, \cite{o2008collisional} predicted a need of 54\% of the crust mass to be ejected in order to account for a 6\% fractionation into the Sm/Nd ratio. This corresponds to about 1.5\% of the Earth's mass that should be ejected as crustal material. This is way lower than our estimates of ejected masses than can go up to 17\% in most erosive cases, even if the fractionation in Sm/Nd predictions remain under the 6\%. For Grand Tack simulations, and if we consider only Earth analogs, the average fraction of ejected mass relative to the final mass of the embryo is 7.5\%$\pm$6.9\%. It should be noted here that \cite{o2008collisional} assume a single impact on a potential Earth analog that would have its current mass. It implies that the effect of accreted material through time (possibly reducing the enrichment of the crust and BSE in Nd relative to Sm during accretion) is fully neglected. Additionally, melting and re-equilibration processes during accretion are not mention in this model neither \citep{o2008collisional}. These different results cannot be directly compared to each other. \newline
Using our results, we also evaluate the hypothesis of enstatite chondrites as the dominant accreting material for Earth as they share strictly similar isotopic compositions with Earth \citep{javoy1995integral, javoy2010chemical}. In this case, collisional erosion has been proposed to account for the elevated Mg/Si ratio of the BSE compared to enstatite chondrites \citep{boujibar2015cosmochemical} if 15\% of the early Earth mass is eroded \citep{boujibar2015cosmochemical}. Only 9\% of the total planetary mass was predicted to be eroded from recent numerical simulations \citep{carter2018collisional} neglecting the impacts of small planetesimals. Our results show that an erosion of 17\% of the final embryo's mass can be reached. Furthermore, this 15\% estimation \citep{boujibar2015cosmochemical} is based on considering a single impact stripping away that amount of a crust already highly enriched in Si compared to Mg (because of the large difference in the degree of compatibility of Si and Mg at the high pressure and temperature conditions considered for differentiation \citep{boujibar2015cosmochemical}). Additionally, the accretion onto its surface of chondritic material is not considered by \cite{boujibar2015cosmochemical} and should lead to a lower fractionation (as illustrated in the present work with the Sm/Nd ratio). Thus, preferential erosion of the crust induced by low energy impacts is unlikely to fully account for the observed Mg/Si ratio of the BSE if Earth was mainly accreted from enstatite chondrites. However, an aternative scenario has been proposed by \cite{dauphas2015planetary} that may explain the Earth Mg/Si non chondritic ratio by nebular fractionation effects, requiring in that case no or low collisional erosion to happen during accretion.\newline
For Grand Tack simulations (but not classical model simulations), we find a positive correlation between the chemical fractionation in Sm and Nd of the BSP and the timing of the last giant impact (fig.~\ref{fig:discussion2}). However there is no correlation between the number of giant impacts and the final fractionation. Thus, the observed correlation between the timing of the Moon-forming event and the fractionation is the consequence of the fact that late last giant impacts are due to a lack of dynamical friction in the disk \citep{jacobson2014lunar}, which consequently creates more erosive collisions. An other key factor for this correlation to the timing of the last giant impact is the crustal growing rate as it grows mostly through the input of chondritic material. Indeed, such amount of chondritic material input further leads to a decrease of the incompatible elements concentration within the crust. This means that for every giant impact a newly more incompatible-rich crust is formed. This makes easier the chemical fractionation for the next impacts responsible for a preferential loss of crust relative to the mantle. Considering chemical measurements on terrestrial and lunar samples (for Hf/W, Mg-suite crustal rocks, highly siderophile elements, I-Xe) \citep{chyba1991terrestrial,walker2009highly, touboul2007late, touboul2009tungsten, touboul2012182w, touboul2015tungsten, pepin2006xenon}, lunar rocks dating \citep{norman2003chronology, nyquist2006feldspathic, borg2011chronological, boyet2015sm, carlson2014rb} and numerical simulations \citep{jacobson2014highly,jacobson2014lunar}, estimates of the age of the moon support a late forming giant impact ranging in between 50 Myr and 150 Myr (after the CAIs formation), with a preferred value around 100 Myr \citep{carlson2014rb, touboul2012182w, pepin2006xenon, jacobson2014lunar}. In addition, two recent studies have argued for a younger estimate of the Moon time formation that would be 50 Myr after CAIs \citep{barboni2017early, thiemens2019early}. These are based on methods (from Lu-Hf system or Hf-W) that have the advantage of presenting relatively low uncertainties. In addition, U-Pb estimates for the age of the terrestrial silicate differentiation is 4480$\pm$20 Myr \citep[e.g.][]{manhes1979lead, albarede1984unscrambling, allegre2008major}, suggesting that the Moon-forming impact is unlikely to have occured after 80Myrs. The age of the Moon is still debated, however most estimates point out to a no younger than 50 Myr Moon-formation. In our Grand Tack simulations the superchondritic Sm/Nd ratio of the BSE can be produced only in the case of a last giant impact occurring after 60 Myrs in agreement with most of these recent estimates. While only a modest fraction (2\%) of our sample of Grand Tack simulations have Earth analogs that suffer a last giant impact after 100 Myr, those accretion histories are favored for another reason. Planetesimals that collide with Earth after the last giant impact deliver highly-siderophile elements (HSEs) to Earth's crust and mantle; matching the very low abundance of HSEs in the BSE \citep{walker2009highly} requires a late last giant impact \citep{jacobson2014highly}. However, after 103 Myr half of the $^{146}$Sm should have decayed into $^{142}$Nd already. This implies that further fractionation between elemental Sm and Nd should have no (or low) effect on the $^{142}$Nd/$^{144}$Nd. In that case, Sm/Nd ratio evolution cannot be a proxy for the  $^{142}$Nd/ $^{144}$Nd after a few tens of Myr and should only be interpreted as an elemental fractionation. Additionally, if an excess in Sm/Nd ratio is produced after 30 Myr, it would cause an anomaly in $^{143}$Nd/$^{144}$Nd larger than observed \citep{boyet2005142nd}. The addition of the evidences for nucleosynthetic anomalies as responsibles for the excess in $^{142}$Nd/$^{144}$Nd and the observables on $^{143}$Nd/$^{144}$Nd ratio lead us to favor a low erosive process during terrestrial planets accretion (i.e. no late last giant impact). However, it is important to mention here the fact that no erosion is assumed in our model during giant embryo-embryo collisions. This has been done by \cite{carter2018collisional} for the 20 first Myr of evolution of the proto-planetary disk. They find fairly similar results than we do, implying that if the results of both studies are applicable together, then the combined effect of larger impacts and cratering impacts could be sufficient to explain a high Sm/Nd ratio acquired during the 30 first Myr. This adds an other argument to the fact that a low erosive environment should be favored for the accretion process not to produce higher anomalies than any observed. Additionally, as mentioned before, the partial melting rate assumed in our present model is low. A simple way to decrease the fractionation in Sm and Nd even for highly erosive cases could be to significantly increase (by several percents at least) the partial melting rate when forming new crusts. The effects of the partial melting rate are presented in the supplementary materials. \newline
Nucleosynthetic anomalies has been emphasized as an explanation for the excess in $^{142}$Nd of the Earth  \citep{burkhardt2016nucleosynthetic, bouvier2016primitive}. That case would require at most 2\% of chemical fractionation of the Sm/Nd ratio induced by collisional erosion in order to maintain the initially implanted Sm/Nd ratio in planetary bodies by nucleosynthetic anomalies  \citep{burkhardt2016nucleosynthetic, bouvier2016primitive}. Therefore, the dynamical and geochemical scenarios leading to such a low final fractionation need to be explored and might provide another set of constraints on the initial conditions of the solar system. \newline 
Impacts with a high velocity (higher than 20 km/s at encounter) are highly erosive \citep{svetsov2011cratering, shuvalov2009atmospheric}. However, they represent only a small fraction of the total. Overall, the impact velocity averaged over 200 Myrs and over the whole disk is higher for non Grand Tack scenarios and for a steeper disc surface densities. Additionally, since the embryos grow faster in the frame of the Grand Tack than in the classical scenario, is it made earlier harder to eject material. As a consequence of these two aspects, the Grand Tack scenarios generally induce a slightly lower fractionation among the different embryos, but the timing of the onset of the Grand Tack might also influence the fractionation yields. Erosion may be more efficient when the system is excited early as embryos have smaller sizes than in a later case. Therefore, future works could constrain the timing of the onset of the Grand Tack (if it existed) in order to preserve the original Sm/Nd fractionation implanted by nucleosynthetic anomalies or, in opposition, to induce efficient stripping of the crust. \newline
Modeling the effects of collisional erosion for other RLE will eventually provide further insights on whether a given terrestrial accretion history can be compatible with observed nucleosynthetic anomalies in chondritic components. 

 \subsection{Outcomes dependency on the dynamical scenario}	
	
     We emphasized that in the case of the remixing scenario reliable on, the grand tack tends to produce a slightly lower final fractionation for relevant Earth analogs compared to the classical scenario (see tables 1 \& 2). However, in the case of smaller embryos, the range of final $\epsilon$ values is larger in the grand tack dynamical scenario than in the classical dynamical scenario. Collisional erosion and consequent chemical fractionation is more efficient for smaller embryos. Within a Grand Tack dynamical scenario, the embryos reach a Earth-like mass faster (within 20-25Myr) and become accretive faster, limiting sooner the erosive power of collisions. The erosive power of collisions is improved at earlier times in the grand tack dynamical scenario since the system is excited by the migration of the giant planets Jupiter and Saturn. This combined effect is responsible for the differences between the dispersion observed in fig~\ref{fig:discussion} for Classical Grand Tack scenarios. However, besides the higher dispersion for the Grand Tack scenario, the final average fractionation is lower for the Grand Tack than for the classical scenario. This is a surprising result that can be explained by the fact that in the Grand Tack scenario the embryos grow much faster. After 20 Myr the embryos ending as good Earth analogs have already reached most of their total masses. From that point, it becomes complicated to produce enough erosion to induce significant fractionation (because of (1) the composition of the crust that is most of the time close to chondritic because of accretion and of (2) the escape velocity that tends to limitate the erosive power of a given impact).

\section{Conclusion}
    
    We have quantified the efficiency of collisional erosion during Earth accretion for two dynamical scenarios ( classical scenario and Grand Tack scenario) and computed the resulting BSE Sm/Nd ratio using a combination of analytical modeling of cratering processes \citep{housen2011ejecta, svetsov2011cratering}, simple mass balance estimates \citep{o2008collisional} and N-body numerical simulations \citep{raymond2009building, jacobson2014lunar}. It is shown (Fig.~\ref{fig:epsilon_evol_fig1}) that only a small amount of impacts (compared to the total) is responsible for a significant fractionation of the BSP Sm/Nd ratio, notably for the classical scenario. During the Grand Tack scenario, most of the fractionation is acquired during the first Myrs (owing to the low escape velocities of small embryos and the rapid decrease of the planetesimal population) of the system evolution while the classical scenario presents a more progressive   fractionation trend, spreading all over the accretion history. We find, that a maximum of  17\% of the proto-planet's mass erosion can be reached for the most erosive cases. In average, over all surviving embryos, only 5\% of the proto-planet's mass is ejected as crust. Even though this amount of ejected crust is large, we show that neither the classical model, nor the Grand Tack model can always account, for the $^{142}$Nd possible excess in terrestrial rocks. However, we find a strong correlation between the fractionation in Sm and Nd with the timing of the Moon-forming event in the case of the Grand Tack scenario. Especially, a large $\epsilon$ fractionation (about 5\%) is reachable for a Moon-forming impact (last-giant impact) occurring after 50 Myr after CAIs, that could account for the proposed off-set in $^{142}$Nd. New evidences of nucleosynthetic heterogeneities in primitive planetary components obviates the need for a superchondritic Sm/Nd ratio of the BSE  and impose limited loss of crust preferential erosion during accretion. Such observations seem to challenge the most eroding scenario explored in this work corresponding to a Grand Tack combined with a late moon forming impact. However, it is important to keep in mind the assumptions of the model presented here. Notably, a low fractionation could be achieved even in an erosive dynamical scenario if the partial melting rate for forming the crust is much larger or under alternative crust compositions as suggested by \cite{carter2018collisional}. Further works should investigate systematically the effect of collisional erosion on the budget of other refractory and lithophile elements and provide an acceptable set of conditions for the dynamical accretion of the Earth that could account for the observed compositions of these elements in the BSE.  
	
	\begin{table}
			\begin{center}
				\renewcommand{\arraystretch}{1.8}
				\resizebox{\textwidth}{!}{
					\begin{tabular}{|c|m{2.3cm}||m{2.6cm}|m{2.6cm}|m{2.6cm}|m{2.6cm}|m{2.6cm}|m{2.6cm}|m{2.6cm}|m{2.6cm}|}
						\hline 
						\multicolumn{2}{|m{2.6cm}||}{ }  & {\Large \textcolor{gray}{cjs1}} & {\Large \textcolor{gray}{cjs15}}  & {\Large \textcolor{gray}{cjsecc15}} & {\Large \textcolor{gray}{jsres}} & {\Large \textcolor{gray}{jsresecc}} & {\Large \textcolor{blue}{eejs15}} & {\Large \textcolor{blue}{ejs1}} & {\Large \textcolor{blue}{ejs15}} \\
						\hline 
						\multirow{2}{.5cm}{(1)} & {\Large {\bf $\bar{\epsilon}$ mean fractionation}} & {\Large 0.029} &  {\Large 0.04}  &  {\Large 0.04} & {\Large 0.041} & {\Large 0.05} & {\Large 0.05} & {\Large 0.04} &  {\Large 0.05} \\
						& {\Large {\bf Standard deviation}}  & {\Large 0.003} & {\Large 0.01} & {\Large 0.01} & {\Large 0.009} & {\Large 0.02} & {\Large 0.015} & {\Large 0.016} & {\Large 0.02} \\
						\hline
						\multirow{2}{.5cm}{(2)} & {\Large {\bf $\bar{\epsilon}$ mean fractionation}} & {\Large 0.007} &  {\Large 0.0049}  &  {\Large 0.004} & {\Large 0.0037} & {\Large 0.004} & {\Large 0.005} & {\Large 0.009} &  {\Large 0.009} \\
						& {\Large{\bf Standard deviation}}  & {\Large 0.002} & {\Large 0.0008} & {\Large 0.001} & {\Large 0.0003} & {\Large 0.0015} & {\Large 0.002} & {\Large 0.006} & {\Large 0.006} \\
						\hline
						\multirow{2}{.5cm}{(3)} & {\Large{\bf $\bar{\epsilon}$ mean fractionation}} & {\Large 0.020} &  {\Large 0.025}  &  {\Large 0.021} & {\Large 0.025} & {\Large 0.030} & {\Large 0.028} & {\Large 0.03} &  {\Large 0.03} \\
						&{\Large {\bf Standard deviation}}  & {\Large 0.002} & {\Large 0.004} & {\Large 0.006} & {\Large 0.005} & {\Large 0.008} & {\Large 0.008}	& {\Large 0.01} & {\Large 0.01} \\
						\hline 
						\multicolumn{2}{|c||}{  {\Large{\bf Grand tack ?}}}  & {\Large no} & {\Large no} & {\Large no} & {\Large no} & {\Large no} & {\Large no} & {\Large no} & {\Large no}  \\
						\hline 
						\multicolumn{2}{|c||} {\Large{\bf Description}} & {\Large Circular Jupiter and Saturn. Mutual inclination = 0.5$^\circ$. $a_j$=5.45AU and $a_s$=8.18AU. $\chi=1$} & {\Large Same as cjs1 but with $\chi=3/2$ (MMSN model)} & {\Large cjs with eccentric orbits. Jupiter and Saturn placed at their cjs semimajor axis 5.45AU and 8.18AU with $e_j$=0.02 and $e_s$=0.03 and a mutual inclination of 0.5$^\circ$} & {\Large Jupiter and Saturn in resonance, placed in their mutual 3:2 mean motion resonance. $a_j$=5.43AU, $a_s$=7.30AU and $e_j$=0.005, $e_s$=0.01 with a mutual inclination of 0.2$^\circ$} & {\Large Jupiter and Saturn in resonance on eccentric orbits. Same as jsres but with $e_s=e_j$=0.03} & {\Large Extra eccentric Jupiter and Saturn. Placed at their actual semi major axis but with high eccentricity orbits. $a_j$=5.25AU, $a_s$=9.54AU, $e_j$=0.1 and $e_s$=0.1. mutual inclination is 1.5$^\circ$} & {\Large Eccentric Jupiter and Satur. Initial orbits around actuals. Same semi major axis and mutual inclination than in eejs15. $e_j$=0.05, $e_s$=0.06} & {\Large Same as ejs1 but with $\chi=3/2$ instead of 1.}  \\
						\hline
				\end{tabular}}
				\caption{Table of the $\epsilon$ mean values in the BSP after the growth of the embryos considered as Earth analogs (with a final mass comprised between 1/2$M_\oplus$ and 1.2$M_\oplus$ and a semi major axis comprised between 0.5 AU and 2 AU) in a classical accretion scenario. The standard deviation associated to each value is also available in the table. (1) refers to the mixing scenario 1, for wich a full reequilibration between crust and mantle is assumed after an impact. (2) denotes the opposite scenario for which there is no reequilibration at all, the giant impact are thus ignored and for the small impacts, it is assumed that everything happens in the crust, remaining the mantle intact. Finally, (3) refers to the intermediate scenario. Giant impacts are treated in agreement with scenario (1) and small impacts are treated the same way than in scenario (2). The values are given for each kind of simulations, with different initial conditions. A quick description of each type of N-body simulation is given. $a_j$ and $a_s$ refer respectively to the Jupiter and Saturn semi-major axis. $e_j$ and $e_S$ denote the eccentricities of Jupiter and Saturn respectively. $\chi$ is used to varie the effect of the disc surface density. The disc surface density can be expressed as: $\sum (r) = \sum_{1}\left(\frac{r}{1AU}\right)^{-\chi}$. The $r^{−1}$ simulations formed slightly fewer planets, contained less total mass in planets and had longer formation timescales for Earth  for the final systems than the $ r^{−3/2}$ simulations\cite{raymond2009building}. "MMSN" refers to the "Minimum Mass Solar Nebula" model\cite{weidenschilling1977distribution}.}
				\label{tab:resultsbysimu}
			\end{center}
		\end{table}

	    \begin{table}
		\begin{center}
			\renewcommand{\arraystretch}{1.8}
			\resizebox{\textwidth}{!}{
					\begin{tabular}{|c|m{5.cm}||m{2.6cm}|m{2.6cm}|m{2.6cm}|}
					\hline 
					\multicolumn{2}{|m{2.6cm}||}{ }  & { \textcolor{gray}{0.025}} & { \textcolor{gray}{0.05}}  & { \textcolor{gray}{0.08}} \\
					\hline 
					\multirow{2}{1.cm}{(1)} &  {\bf $\bar{\epsilon}$ mean fractionation} &  0.11 &  0.11  &   0.13   \\
					&  {\bf Standard deviation}  &  0.04 &  0.02 &  0.03 \\
					\hline
					\multirow{2}{1.cm}{(2)} & {\bf $\bar{\epsilon}$ mean fractionation} & 0.0019 &  0.0021  &   0.0007 \\
					& {\bf Standard deviation} & 0.0007 &  0.0005 &  0.0002 \\
					\hline
					\multirow{2}{1.cm}{(3)} & {\bf $\bar{\epsilon}$ mean fractionation} & 0.03 &   0.021  &  0.013 \\
					&{\bf Standard deviation} & 0.01 &  0.007 &  0.008 \\
					\hline 
					\multicolumn{2}{|c||}{  {\bf Grand tack ?}}  & yes &  yes &  yes   \\
					\hline 
					\multicolumn{2}{|c||} {{\bf Description}} &  initial embryos masses  = 0.025$M_{\oplus}$.   &  initial embryos masses  = 0.05$M_{\oplus}$ & initial embryos masses  = 0.08$M_{\oplus}$  \\
					\hline
				\end{tabular}}
			\caption{Table of the $\epsilon$ mean values in the BSP after the growth of the embryos considered as good Earth analogs in a grand tack accretion scenario \citep{jacobson2014lunar}. The standard deviation associated to each value is also available in the table. (1) refers to the mixing scenario 1, for wich a full reequilibration between crust and mantle is assumed after an impact. (2) denotes the opposite scenario for which there is no reequilibration at all, the giant impact are thus ignored and for the small impacts, it is assumed that everything happens in the crust, remaining the mantle intact. Finally, (3) refers to the intermediate scenario. Giant impacts are treated in agreement with scenario (1) and small impacts are treated the same way than in scenario (2). The values are given for the three different sets of initial conditions concerning the initial mass of an embryo (corresponding to the mass that basically has an embryo when the inward migration of Jupiter begins). $M_{\oplus}$ refers to the actual Earth's mass.} 
			\label{tab:resultsbysimu2}
		\end{center}
	\end{table}

\section*{Acknowledgement}
J.S. acknowledges support from the French National Research Agency (ANR project VolTerre, grant no. ANR- 14-CE33-0017-01) And Institut Universitaire de France.
Parts of this work were supported by the UnivEarthS Labex programme at Sorbonne Paris Cit\'e (ANR-10-LABX-0023 and ANR-11- IDEX-0005-02).

	\nolinenumbers
	\newpage
	\bibliographystyle{naturemag}
	\bibliography{biblio}

\section*{Appendice}
	\label{sec:annexes}
	
	\subsection{Influence of the partial melting rate}

\begin{figure}
		\centerline{\includegraphics[width=.7\linewidth]{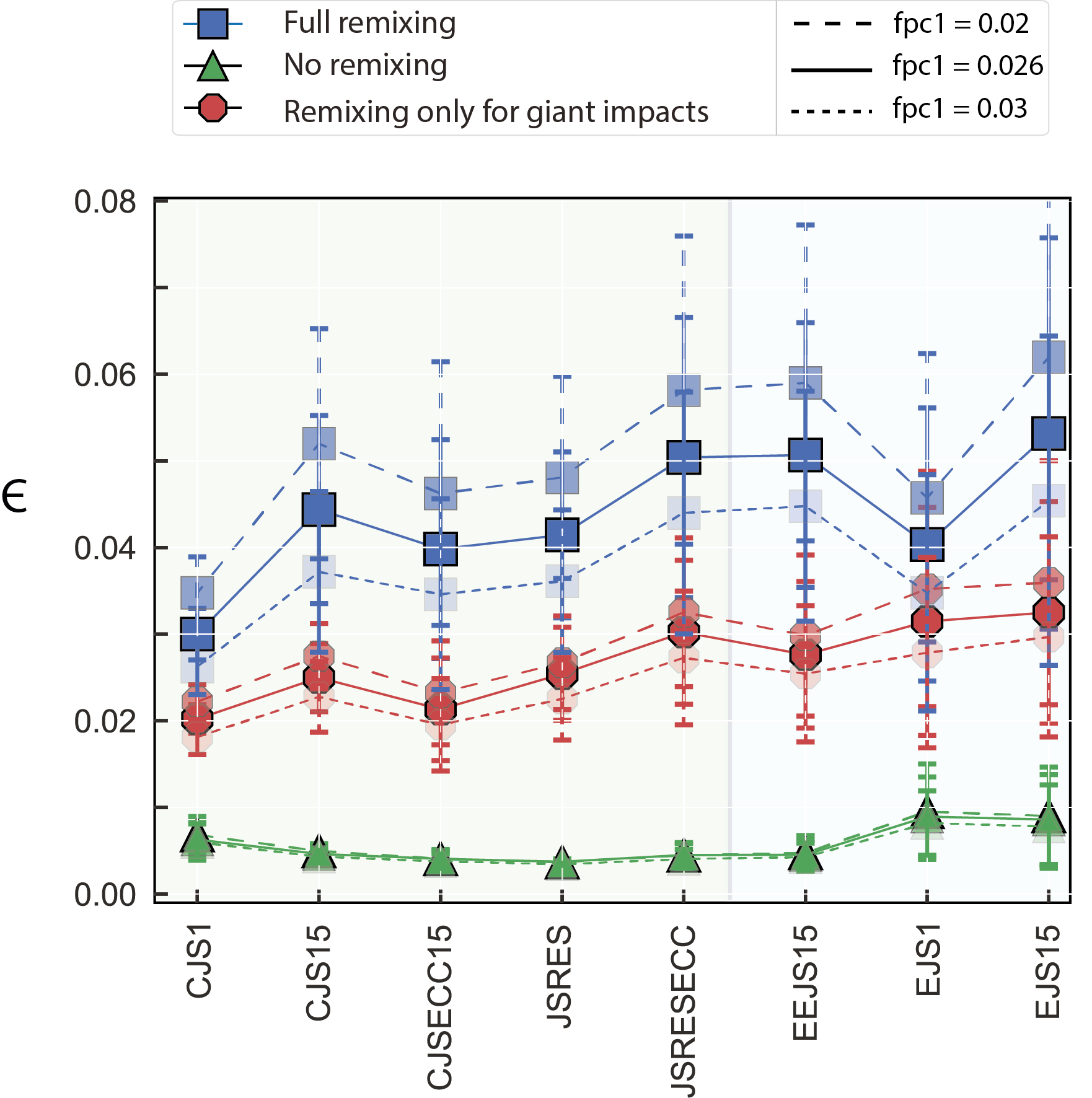}}
		\caption{Partial melting rate influence on the final embryos fractionnation evidenced by representation of $\epsilon$ mean values for the different dynamical scenarios in a classical accretion disk. Solid lines represent the model presented in the main text while dashed lines represent the results for other $\alpha$ tested here. The colors and markers refer to the mixing scenarios.}
		\label{fig:partial_melting}
	\end{figure}

When forming a new crust, the mantle partial melting rate is determinant for the repartition of the chemical elements between crust and mantle. As it is evidenced in the eq.\ref{eq:MBdiff}, the higher the partial melting rate $f_{p-c}^1$, the lower the concentration in Nd within the crust. As a consequence, the preferential loss of Nd compared to Sm is lower for a higher partial melting rate. A given amount of eroded crust leads to a lower fractionation in the Sm/Nd ratio for a higher $f_{p-c}^1$. The results for three different partial melting rates are presented in fig.\ref{fig:partial_melting}. This parameter clearly has a non neglectable influence on the final results; however, since when it increases the fractionation decreases. Accordingly, $f_{p-c}^1$ has been chosen low (2.6\%) to estimate an upper bound of possible impact-induced fractionation. 

\subsection{Influence of the Size-frequency distribution}
	
\begin{figure}
		\centerline{\includegraphics[width=0.7\linewidth]{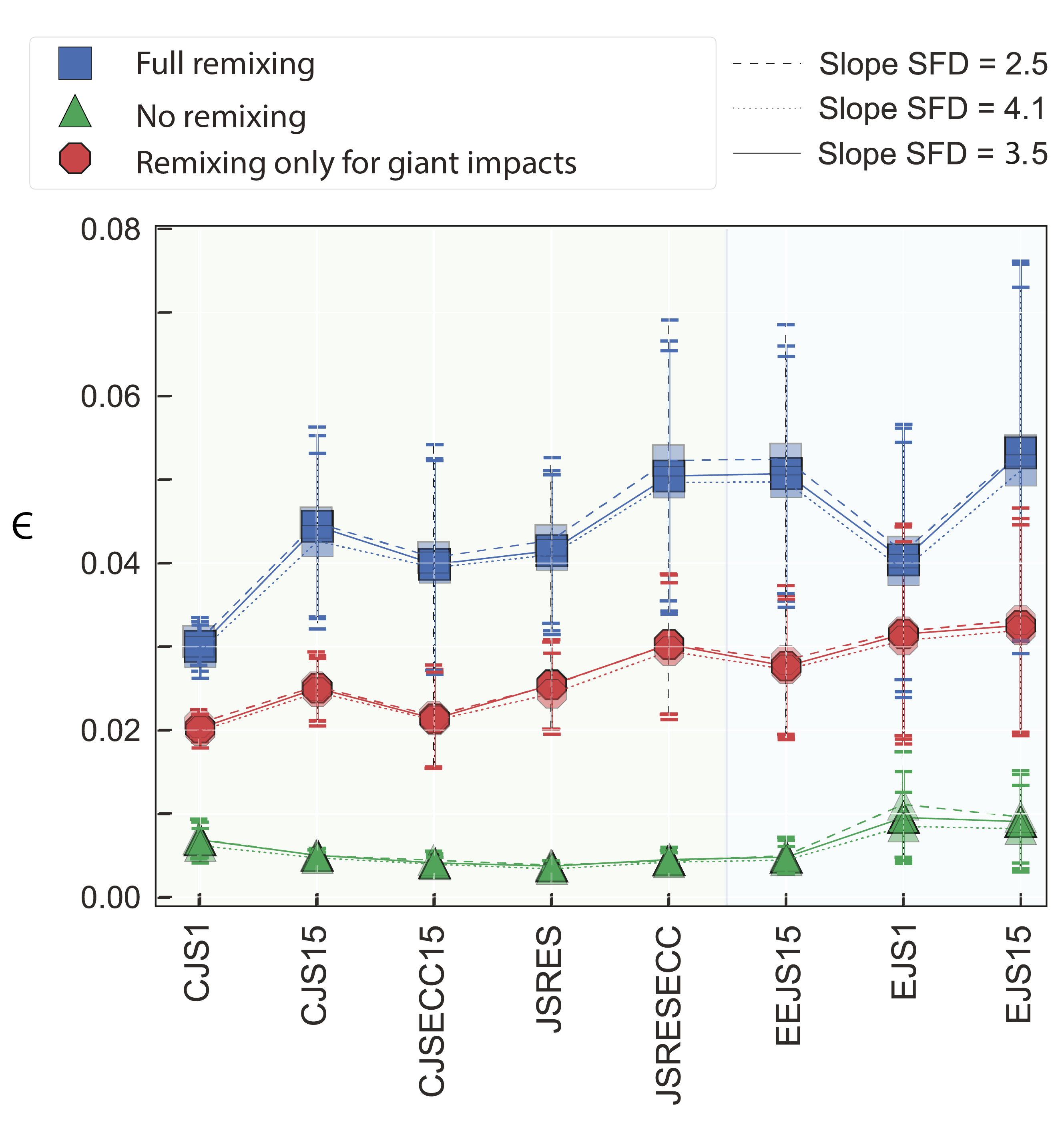}} 
		\caption{$\alpha$ (SFD slope) influence on the final embryos fractionnation evidenced by representation of $\epsilon$ mean values for the different dynamical scenarios. Solid lines represent the model presented in the main text while dashed lines represent the results for other $\alpha$ tested here. The colors and markers refer to the mixing scenarios.}
		\label{fig:slope_influence}
	\end{figure}

	The final \label{sec_an:slope}results concerning the evolution of the fractionation value ($\epsilon$) over time have been computed for 3 other $\alpha$: 2.5, 3 and 4.1. As the concern is only the influence of the slope over the fractionation, we present it only for the different classical simulations, that in average, produce a higher fractionation value than the Grand Tack scenario. They are presented in fig.~\ref{fig:slope_influence}. The solid colored lines represent the results presented above (with $\alpha=3.5$), the dashed lines represents the results of the three different mixing scenarios for $\alpha=2.5$ and the last represents the results for $\alpha=4.1$. The higher $\alpha$ (i.e. more mass in the smaller bodies), the lower $\epsilon$. It may be expected that earlier in the planetary formation history, the slope would have been slightly higher \citep{tanaka1996steady}. As a consequence, the fractionation estimated in this study could be overestimated. However, the difference between the different models from different $\alpha$ are very low, so the choice of this parameter is not critical. 
	
	\subsection{Correlation with the last giant impact is not due to the number of giant impacts}
	
\begin{figure}
		\centerline{\includegraphics[width=1.\linewidth]{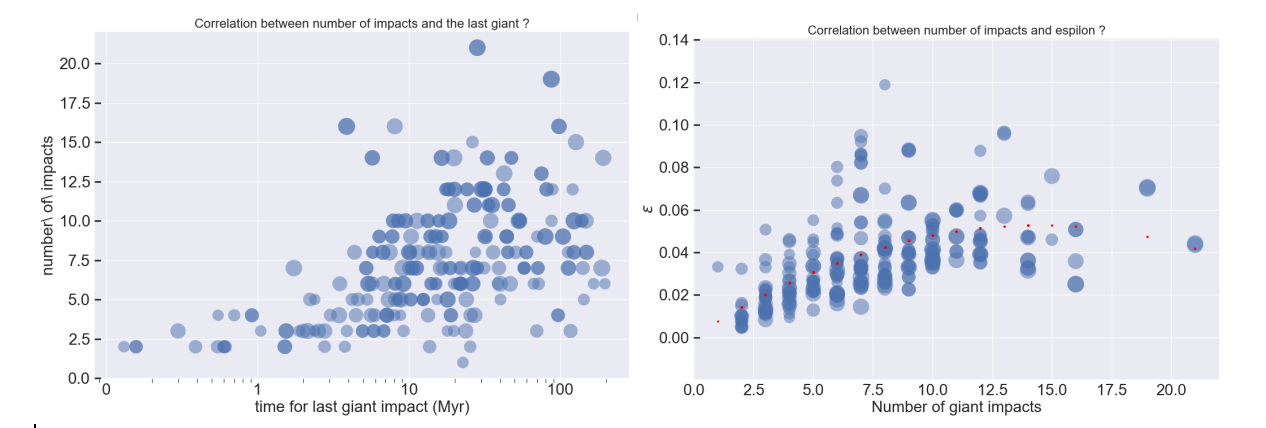}} 
		\caption{(1) Left panel.  Number of impacts suffered by the embryos surviving to Grand Tack simulations as a function of the timing for the last giant impact. 
		(2) Right panel. $\epsilon$ value for epsilon as a function of the number of giant impacts for Grand-Tack surviving embryos. 
		In both figures the sizes of the circles stand the final masses of the surviving embryo.}
		\label{fig:class_llg}
	\end{figure}

	Figure~\ref{fig:class_llg} shows 1) on the left panel the number of impacts suffered by the embryos surviving to Grand Tack simulations as a function of the timing for the last giant impact; and 2) on the right panel, the epsilon value for epsilon as a function of the number of giant impacts for Grand-Tack surviving embryos. We see no correlation between any of these values. The only thing that can be pointed out by this figure is the dispersion after a few Myr that becomes extremely large. The denotes the fact that the correlation observed for the Grand Tack scenario betweem the fractionation of the Earth analogs and the time they registered their last involvement into a giant impact is not linked to the number of giant impacts they went through. The correlation is a correlation to the timing of the last chemical re-equilibration and not on how often a re-equilibration has occured. 

	\subsection{Influence of the choice for the cratering model}
	The scaling laws for rocky targets \citep{holsapple2007crater} has been chosen when integrated over all impact angles \citep{svetsov2011cratering} for the description of the eroded mass. However, \cite{svetsov2011cratering} explores and improves an other scaling law that has been suggested \citep{shuvalov2009atmospheric} based on simulations of impacts in the presence of planetary atmospheres. This very last description is tested in this study, to explore the influence that could have the choice of the cratering model used. This formulation of eroded mass is described before discussing the associated results.\newline
	The escaped mass is estimated as \cite{svetsov2011cratering}:
		\begin{equation}
	   				M_{escaped}=m_{imp}C(V_{imp})\frac{V_{imp}^2-C_1(V_{imp})v_{esc}^2}{v_{esc}^2}\left(\frac{V_{imp}}{20v_{esc}}\right)^a,
	   				\label{eq:mesc}
	   			\end{equation}	
	with $m_{imp}$ and $V_{imp}$ being respectively the impactor mass and the impact velocity. Here, $v_{esc}$ is assumed to be the two-body escape speed:
	
	\begin{equation}
		v_{esc}=\sqrt{\left(\frac{2G(m_{tar}+m_{imp})}{R_{tar}+R_{imp}}\right)},
		\label{eq:vesc}
	\end{equation}
	with $m_{tar}$ the target mass, $R_{tar}$ and $R_{imp}$ are respectively  the target and the impactor radius. The different dimensionless coefficients needed for the calculation of $M_{escaped}$, are defined as follows: 
		
	\begin{equation}
	C_1(V_{imp})=0.7+(V_{imp}/20km/s)^2, 
	\end{equation}
	\begin{equation}
	a=0.15-0.0003(V_{imp}-15km/s)^2,
	\end{equation}	
	And, concerning C (wich is also a dimensionless number):	
	\begin{itemize}
		\item if $V_{imp} \geq 20\ km/s$, then $C(V_{imp})=0.02$,
		\item if $V_{imp} = 10\ km/s$, then $C(V_{imp})=0.035$,
		\item if $V_{imp} \leq 5\ km/s$, then $C(V_{imp})=0.07$,
		\item if $5\ km/s < V_{imp} < 10\ km/s$, then $C(V_{imp})=0.07+[(0.035-0.07)/5](V_{imp}-5)$,
		\item if $10\ km/s < V_{imp} < 20\ km/s$, then $C(V_{imp})=0.02+[(0.02-0.035)/10(V_{imp}-10)$.
	\end{itemize}
	The two different scaling laws are tested over all the collisional histories of the different embryos issued from the numerical simulations. Depending on the range of impact velocity, either one model overestimates the escaped mass compared to the other, either it underestimates it. However, the difference between the two different models is not critical. For both models the final $\epsilon$ values are similar within a few percents and show a $\epsilon$ below 0.05.	
	   
	\subsection{Three sets of assumptions for mass balance modeling}
	    
	    We present here the description of the two additional mixing scenarios for chemical mass balance resulting from impacts. These are end-members meant to provide lower-bound and maximum-bound fractionation.
	    
	    \subsubsection{Scenario 1:  "the full remixing" model}
		
		The BSP is molten and the newly formed crust reequilibrates with the mantle after each impact. Giant impacts are considered as perfect mergers, and a full reequilibration between crust and mantle is still assumed. Expressions~\ref{eq:MBdiff} and~\ref{eq:MBerod} are used for estimating the resulting BSP concentration in a given element $M$ compared to its previous value. 
		
		\subsubsection{Scenario 2: "no remixing" model}
		
		 Only small impacts are taken into account for estimating the masses of eroded crust and accreted chondritic material. All accreted material is assumed to fall onto the surface so that only the crustal composition is modified while mantle composition (and its mass) remains constant. No new crust is formed. Accordingly, the new composition of the crust in $M$ at a given time $i$ is expressed following:
		
			\begin{equation}
		        C_{Mcrust}^i = \frac{C_{Mcrust}^{i-1}m_{crust}^{i-1}+C_M^{chondritic}m_{accreted}^{i} - C_{Mcrust}^{i-1}m_{ejected}^{i}}{m_{crust}^{i-1}+m_{accreted}^{i}-m_{ejected}^{i}},
		        \label{eq:MBcrust}
        	\end{equation}
        where $m_{crust}$ is the crustal mass, $C_M^{chondritic}$ is the chondritic concentration for the element $M$, $m_{accreted}$ is the accreted mass (from impactor) and $m_{ejected}$ is the escaped mass. 
        	
		In this scenario, the giant impacts are not treated at all  in terms of mass balance. It means that when a giant impact is recorded into N-body collisions file, our model is ignoring it, it behaves exaclty as if proto-Earth remained perfectly intact. No new crust is formed from partial melting. It is far from realistic scenario, but is meant to represent a end-members.
	    
\begin{figure}
		\centerline{\includegraphics[width=1.\linewidth]{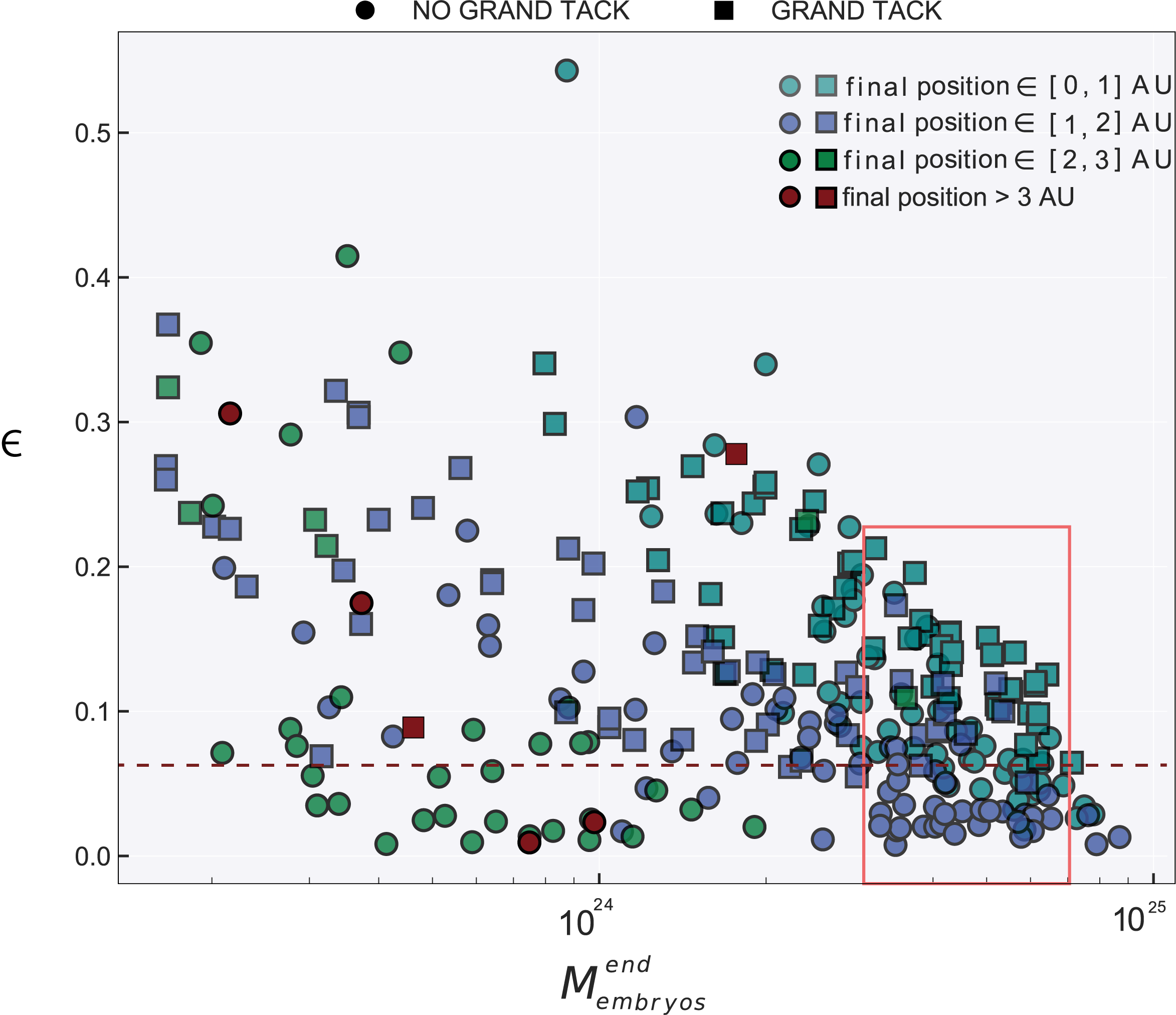}}
		\caption{$\epsilon$ (Sm-Nd fractionation) with respect to chondritic Sm/Nd values of the different surviving embryos at the end of classical and Grand Tack N-body simulations reported as a function of their final masses. The remixing model adopted here is the most favorable to fractionation (scenario 1: full remixing). The markers refer to the type of accretionary scenario (circles for the classical scenario and square for grand tack). The colors refer to the final semi-major axis of the embryos. The red box shows the Earth analogs.}
		\label{fig:eps_s1}
	\end{figure}

\begin{figure}
		\centerline{\includegraphics[width=1.\linewidth]{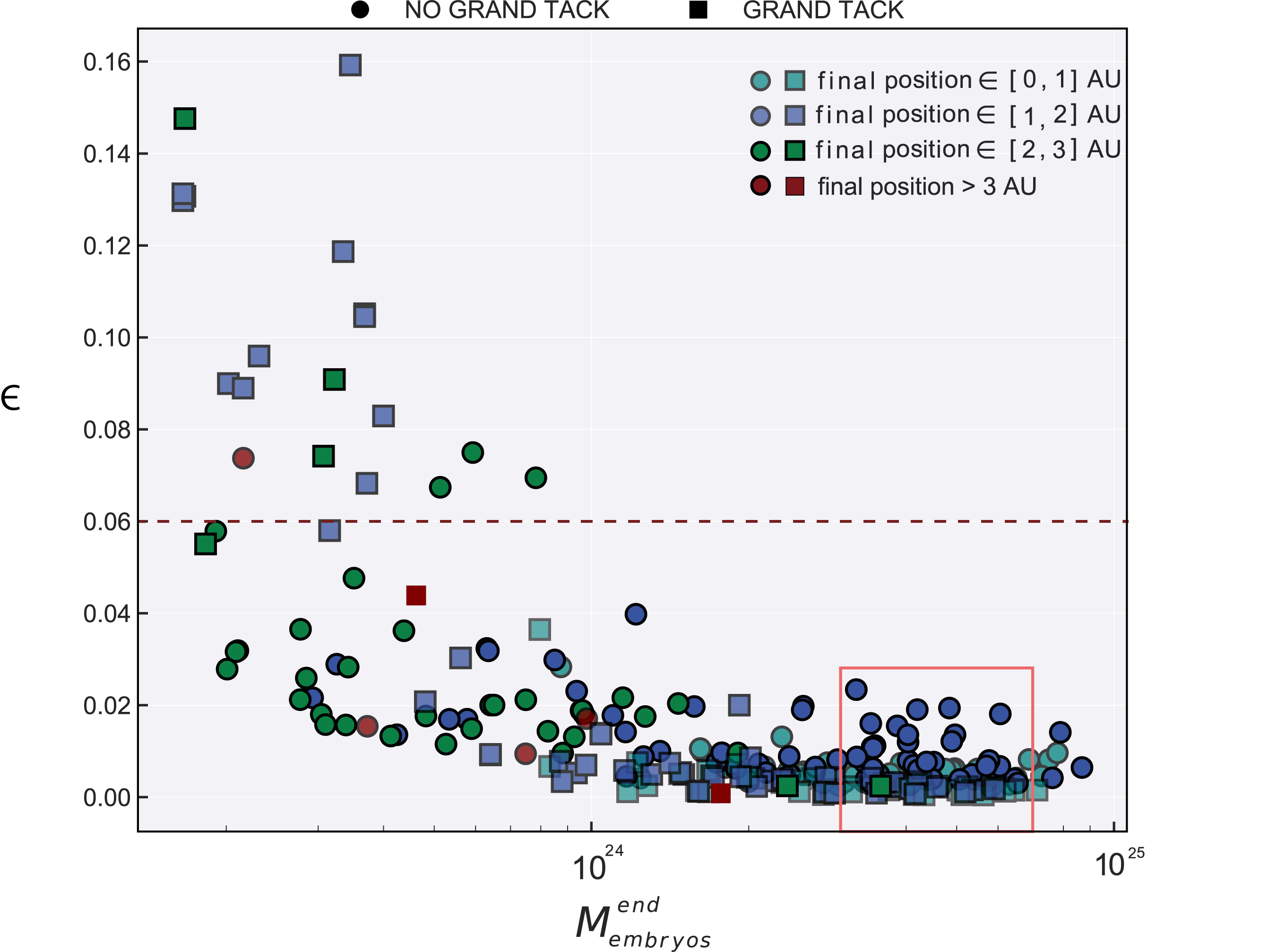}}
		\caption{$\epsilon$ (Sm-Nd fractionation) with respect to chondritic Sm/Nd values of the different surviving embryos at the end of classical and Grand Tack N-body simulations reported as a function of their final masses. The remixing model adopted here is the less favorable to fractionation (scenario 2: no remixing). The markers refer to the type of accretionary scenario (circles for the classical scenario and square for grand tack). The colors refer to the final semi-major axis of the embryos. The red box shows the Earth analogs.}
		\label{fig:eps_s2}
	\end{figure}

	    The scenario \label{sec:3_sets} 1 (full remixing) and 2 (no remixing) represent unrealistic end members. As a consequence, previously, only the most realistic case, the scenario 3 (remixing only for giant impacts) has been presented. The results concerning the other two cases are presented here in details (\ref{fig:eps_s1} and \ref{fig:eps_s2}). The values of $\epsilon$ range from $0.029\pm0.003$ to $0.05\pm0.02$ for the classical scenario in the case of the mixing scenario (1) (fig. \ref{fig:eps_s1}) which is clearly the most favorable to fractionate the Sm/Nd ratio with respect to chondritic composition. In this case, a new crust is formed after each impact and Sm and Nd are re-distributed according to their partition coefficients between the solid mantle and the liquid crust. This leads to an increase of the Nd/Sm ratio of the newly formed crust at each step of the simulation as Nd is more incompatible than Sm. However, this scenario is not appropriate as low energy events are not likely to produce large amounts of melt and can be consider as an upper bound for the final modeled Sm/Nd ratio of the bulk silicate Earth. In the case of the mixing scenario (2) (fig. \ref{fig:eps_s2}, $\epsilon$ takes values between $0.0037\pm0.0004$ and $0.009\pm0.006$ for a classical accretion scenario. In this scenario the crust is refertilized in chondritic Sm/Nd after each impact and absence of melting leads to the lowest chemical fractionation of the Sm/Nd value. The crust mass increases and becomes more and more chondritic during the course of accretion. This scenario can be considered as the opposite as scenario 1, and provides a lower bound for $\epsilon$ values. The global trand of those observation is found also concerning the grand tack simulations. However, the final fractionation values are much more spread out. Especially, concerning the most favorable case to fractionation: full remixing; the values of the $\epsilon$ fractionation are very high in average ($0.12\pm0.03$). Such fractionation is unexpected as we do not sample any extraterrestrial body or meteorite that would present such values of Sm/Nd ratio. However, since the grand tack has a powerful effect of fractionation during the first Myrs of accretion, is the crust is enriched in Nd at each step because of a new crust-forming event du to complete melting of the silicate Earth, it is clear that the total fractionation will take unrealistically large values. As a conclusion, those two end members gives us clues about the very lower and upper bound of the fractionation, but they still quite unrealistic and they are not strong enough to allow a better approach and constrain the Sm/Nd ratio evolution over accretion histories. That is the reason why the intermediate case (3) - remixing only for giant impacts - has been chosen alone to follow this fractionation. It provides a good first approximation of the fractionation as a response to collisional erosion. 
	
\end{document}